\documentclass[prd,twocolumn,tightenlines,superscriptaddress,nofootinbib,
showpacs]{revtex4}
\usepackage{amssymb,latexsym}
\usepackage{amsmath,amsbsy,bbm}
\usepackage{epsfig,bm}
\usepackage{graphicx,comment}
\usepackage{color}
\usepackage[normalem]{ulem}
\begin{document}

\title{Universal properties of
three-dimensional random-exchange quantum antiferromagnets}

\author{D.-R. Tan}
\affiliation{Department of Physics, National Taiwan Normal University,
88, Sec.4, Ting-Chou Rd., Taipei 116, Taiwan}
\author{F.-J. Jiang}
\email[]{fjjiang@ntnu.edu.tw}
\affiliation{Department of Physics, National Taiwan Normal University,
88, Sec.4, Ting-Chou Rd., Taipei 116, Taiwan}

\begin{abstract}
The thermal and ground state properties of a class of three-dimensional (3D) 
random-exchange spin-1/2 antiferromagnets are studied using first principles 
quantum Monte Carlo method. Our motivation is to examine whether the newly 
discovered universal properties, which connect the N\'eel temperature
and the staggered magnetization density, for the clean 3D quantum dimerized 
Heisenberg models remain valid for the random-exchange models considered here. 
Remarkably, similar to the clean systems, our Monte Carlo results indicate that 
these universal relations also emerge for the considered models with the 
introduced antiferromagnetic randomness. The scope of the validity of these 
universal properties for the 3D quantum antiferromagnets is investigated as 
well. 
\end{abstract}


\maketitle

\section{Introduction}
Heisenberg-type models have been studied extensively using both analytic 
and numerical methods during the last two decades 
\cite{Cha88,Hal88,Reg88,Cha89,Sin89,Sin90,Zhe91,Chu94,Oit94,Tro95,Bea96,San97,Mat02,Wan05,Ng06,Par08,Jia09.1}. 
This is mainly due to the fact that these models can qualitatively or 
quantitatively describe experimental data. In real materials, impurities 
often play important roles in their properties. As a result, 
generalized Heisenberg-type models such as taking into account the effects 
of (non)magnetic impurities have been explored in great detail as well 
\cite{Sac99,Sac00,Sac01,Tro02,Sac03,Sus03,Hog03,Hog04,Hog07,Hog071}. In 
additional to impurities, quench disorder effects, namely randomness
in the strength of the antiferromagnetic coupling, also attract a lot of 
theoretical interest \cite{San95,Mel02,Lin03,Laf06,Lin06}. 
Indeed, investigation associated with impurity and 
disorder effects for antiferromagnets has led to several exotic 
(theoretical) results. Two notable such examples are the anomalous Curie 
constant \cite{Hog07} and violation of Harris criterion \cite{Voj10}. 
While most of these studied are related to 
two-dimensional models, experimental results, 
like those of TlCuCl$_3$\cite{Cav01,Rue03,Rue08}, have triggered 
theoretical attention to higher dimensional systems as well 
\cite{Oit04,Noh05,Kul11,Oit12,Jin12,Kao13}. It should be pointed out
that for three spatial dimensions which is the upper critical dimension, 
the relevant universal physical
quantities close to (second order) quantum phase transitions, namely the 
critical exponents are described by the corresponding 
mean-field values (Beside the leading exponents, there are logarithmic 
corrections to the scaling as well). 
Still, investigating universal relations that 
do not depend on the microscopic details of the systems is an
interesting research topic. Such studies are also practically useful from a
experimental point of view.     
    

Recently, close to quantum critical points (QCPs), several universal relations 
between the N\'eel temperature 
$T_N$ and the staggered magnetization density $M_s$ are established for 
three-dimensional (3D) spin-1/2 dimerized antiferromagnets both theoretically 
and experimentally \cite{Oit12,Jin12,Kao13,Mer14}. For instance,
as functions of $M_s$, the physical quantities $T_N/\overline{J}$
and $T_N/T^{\star}$ show universal dependence on $M_s$ \cite{Jin12}.
Here $\overline{J}$ and $T^{\star}$ are the averaged strength of 
antiferromagnetic couplings and the temperature where peak of the uniform 
susceptibility as a function of temperatures takes place, respectively.
Notice these universal relations simplify to linear ones in the 
vicinity of quantum critical points. Interestingly, for 3D random-exchange 
systems, it is demonstrated in Refs.~\cite{Kao14,Kao14.1} that the long-range 
antiferromagnetism is robust against both the box-like and the singular
disorder distributions. In addition, a linear dependence of 
$\overline{T_N}/\overline{J}$ on $\overline{M_s}$ close to the corresponding 
data points of the clean systems is observed in Refs.~\cite{Kao14,Kao14.1} 
as well (In this study observables with a overline on them stand for the 
disordered average). This is quantitatively different from the scenario of 
clean 3D dimerized Heisenberg models. Notice for the random-exchange models 
considered in Refs.~\cite{Kao14,Kao14.1}, the related quantum phase transitions
do not occur at finite randomness in the relevant parameter spaces. Hence it 
would be extremely interesting to 
investigate whether the universal relations between $T_N$ and $M_s$, which are 
found for the clean 3D quantum dimerized systems, remain valid for 3D spin-1/2 
random-exchange models with the corresponding quantum critical points 
taking finite values in the related parameter spaces. Motivated by this, here we 
investigate a class of 3D quantum 
random-exchange Heisenberg models which undergo quantum phase 
transitions with randomness of finite magnitude.
These models will be introduced explicitly later. 
Remarkably, the universal relations observed in Ref.~\cite{Jin12} are 
valid for the models considered in this study. Specifically, for our models, 
both the quantities $\overline{T_N}/\overline{J}$ and 
$\overline{T_N}/\overline{T^{\star}}$ do show universal dependence on 
$\overline{M_s}$. We have further studied the scope of the validity of these 
universal relations and concluded that such properties are justified within 
each individual category (explained later).
  
The rest of this paper is organized as follows. After the introduction,
the studied models, the employed technical methods, the used randomness, as 
well as the observables are described in section 2. Section 3 contains the 
detailed numerical results and discussion. Finally we concludes our 
investigation in section 4.

\section{Microscopic model, bond randomness and observables}
\begin{figure}
\begin{center}
\vbox{
\includegraphics[width=0.225\textwidth]{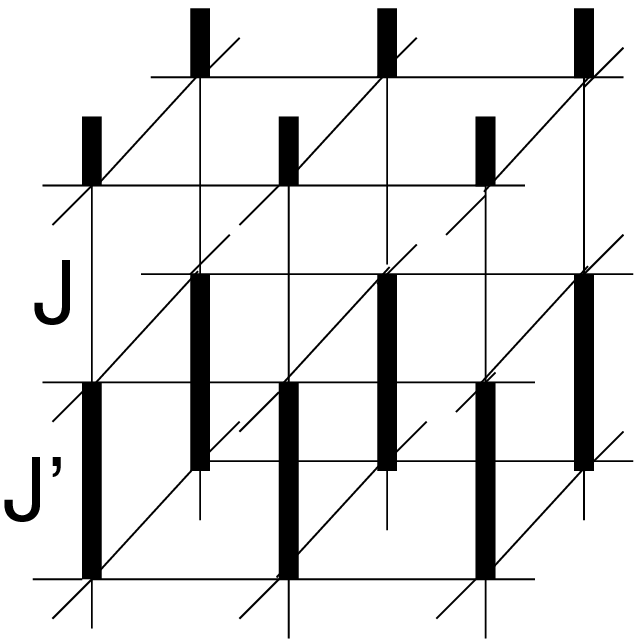}\vskip0.4cm
\includegraphics[width=0.225\textwidth]{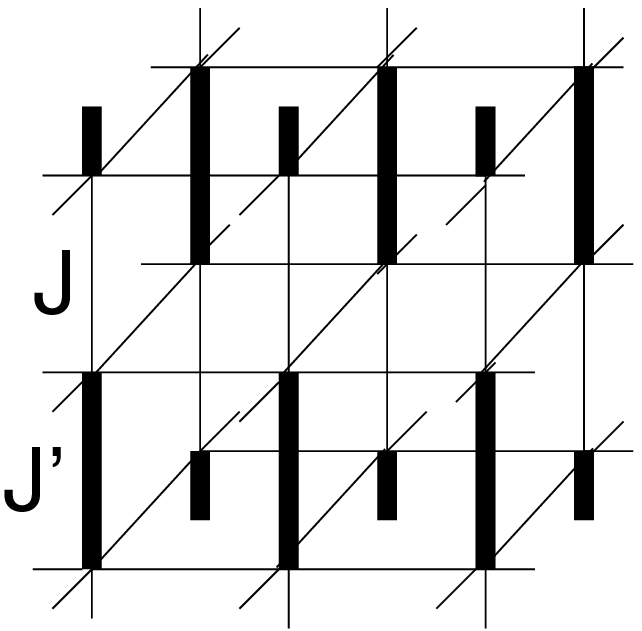}
}
\end{center}\vskip-0.5cm
\caption{The three dimensional random ladder- (top panel) and staggered-dimer
(bottom panel) quantum Heisenberg models considered in this study.}
\label{fig0}
\end{figure}

The studied 3D random-exchange quantum Heisenberg models are given by 
the Hamilton operators
\begin{eqnarray}
\label{hamilton}
H = \sum_{\langle ij \rangle}J_{ij}\,\vec S_i \cdot \vec S_{j} 
+ \sum_{\langle i'j' \rangle}J'_{i'j'}\,\vec S_i \cdot \vec S_{j} ,
\end{eqnarray}
where $J_{ij}$ and $J'_{i'j'}$ are the antiferromagnetic exchange coupling 
connecting nearest neighbor spins $\langle  ij \rangle$ and  
$\langle  i'j' \rangle$, respectively, and $\vec{S}_i$ is
the spin-1/2 operator at site $i$. For simplicity, here we will
call $J_{ij}$ and $J'_{i'j'}$ the (antiferromagnetic) bonds. Fig.~\ref{fig0} 
demonstrates the models described by Eq.~(\ref{hamilton}) in the pictorial 
form. Notice in this study the models shown in the top and bottom panels of
Fig.~\ref{fig0} will be called the (random) ladder- and staggered-dimer 
models, respecitvely. The randomness to the antiferromagnetic coupling 
strength is introduced as follows. First of all,
the strength of the antiferromagnetic couplings $J_{ij}$ are set to 1.0
for both models. Second, $J'_{i',j'}$ takes the value
of $J_c(1.0+D)$ and $J_c(1.0-D)$ randomly with probabilities
$P$ and $1-P$, respectively. Here $0.0< D < 1.0$ and $0.0 < P \le 1.0$.
The $J_c$ appearing above is the critical coupling
of the clean system which is given by $J_c = 4.0128$ ($J_c = 4.6083$)
for the random ladder-dimer (staggered-dimer) quantum Heisenberg 
model \cite{Noh05}. A similar (2D) model has been considered in \cite{San14} as well.  
Based on the method of inducing randomness, there exists a critical
$P_c$ at which a quantum phase transition from the antiferromagnetic long-range
order to a disordered phase occurs. Here we will focus on the 
case of $D=0.5$. To determine the N\'eel temperature 
$\overline{T_N}$ and the staggered magnetization density $\overline{M_s}$ 
for the considered models with the employed randomness, the observables
staggered structure factor $\overline{S(\pi,\pi)}$, spin stiffness $\overline{\rho_s}$, 
and second Binder ratio $\overline{Q_2}$ are calculated in our simulations. 
The quantity $\overline{S(\pi,\pi)}$ is defined by 
\begin{equation}
\overline{S(\pi,\pi)} = 3 \langle ( m_s^z )^2\rangle
\end{equation}
on a finite cubical lattice with linear size $L$. Here 
$m_s^z = \frac{1}{L^3}\sum_{i}(-1)^{i_1+i_2+i_3}S^z_i$ with $S^{z}_i$ being 
the third-component of the spin-1/2 operator $\vec{S}_i$ at site i. 
In addition, $\overline{\rho_s}$ is given as
\begin{equation}
\overline{\rho_s} = \frac{1}{3\beta L^3}\sum_i\langle W_i^2\rangle,\,\,i\in\{1,2,3\} 
,\end{equation}
where $W_i$ is the winding number in $i$ direction and $\beta$ is the inverse
temperature. Finally, the observable $\overline{Q_2}$ takes the form
\begin{equation}
\overline{Q_2} = \frac{\langle (m_s^z)^2 \rangle ^2}{\langle (m_s^z)^4\rangle}.
\end{equation}
With these observables, $\overline{T_N}$ 
and $\overline{M_s}$ can be determined with high precision.

\begin{figure}
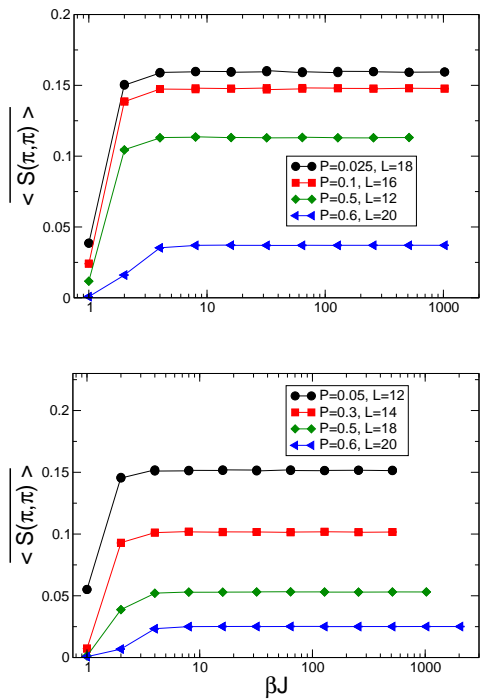

\begin{center}
\vbox{
\includegraphics[width=0.35\textwidth]{ladder_beta_convergence.eps}\vskip0.4cm
\includegraphics[width=0.35\textwidth]{staggered_beta_convergence.eps}
}
\end{center}\vskip-0.5cm
\caption{Convergence of $\overline{S(\pi,\pi)}$ to the ground state values, for 
several considered $P$ and various box sizes $L$, for the random ladder- (top 
panel) and staggered-dimer (bottom panel) models. The solid lines are added to 
guide the eye.}
\label{fig1}
\end{figure}

\section{The numerical results}

To understand whether universal relations between $\overline{T_N}$ 
and $\overline{M_s}$, similar to those of clean dimerized systems, appear 
for the studied 3D quantum Heisenberg models with the introduced bond 
randomness, we 
have carried out a large-scale Monte Carlo simulation using the stochastic 
series expansion (SSE) algorithm with very efficient loop-operator update 
\cite{San99}. In particular, the $\beta$-doubling scheme described in 
\cite{San02} is used in our investigation in order to access the ground 
state properties in an efficient way. Furthermore, each random configuration 
is generated by its own seed and several hundred (for obtaining ground state 
properties) to several thousand (for the determination
of thermal properties) randomness realizations are produced. 
The potential systematic uncertainties due to thermalization, Monte Carlo
sweeps within each randomness realization, as well as 
the number of configurations used for disordered average are examined by 
carrying out many trial simulations. The resulting conclusions based on these 
trial simulations are consistent with those presented in this study.

\subsection{The determination of $\overline{M_s}$}
 
\begin{figure}
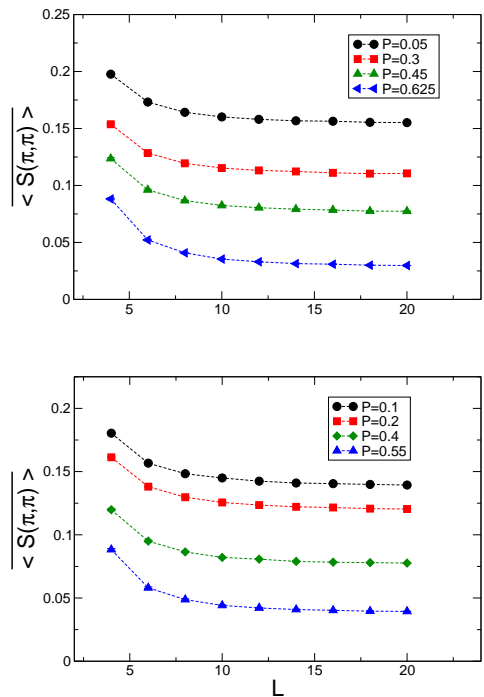

\begin{center}
\vbox{
\includegraphics[width=0.35\textwidth]{ladder_structure_factor_L.eps}\vskip0.5cm
\includegraphics[width=0.35\textwidth]{staggered_structure_factor_L.eps}
}
\end{center}\vskip-0.5cm
\caption{$L$ dependence of the staggered structure factors 
$\overline{S(\pi,\pi)}$, at several considered values of $P$, for both the 
random ladder- (top panel) and staggered-dimer (bottom panel) models.
The dashed lines are added to guide the eye.}
\label{fig2}
\end{figure}

\begin{figure}
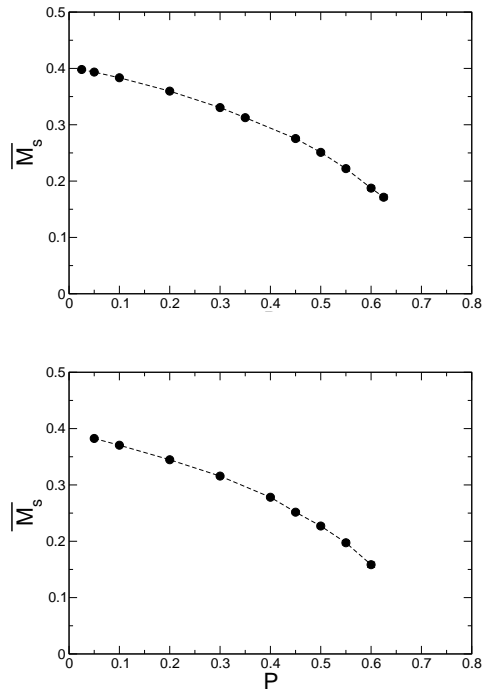

\begin{center}
\vbox{
\includegraphics[width=0.35\textwidth]{ladder_stag_mag_density.eps}\vskip0.45cm
\includegraphics[width=0.35\textwidth]{staggered_stag_mag_density.eps}
}
\end{center}\vskip-0.5cm
\caption{$\overline{M_s}$ as functions of the considered values of 
$P$ for both the studied models. While the top panel is for the random
ladder-dimer model, the results of $\overline{M_s}$ for the random 
staggered-dimer model is depicted in the bottom panel. The dashed lines are 
added to guide the eye.}
\label{fig3}
\end{figure}

The reach of the ground state values of  
$\overline{S(\pi,\pi)}$ for several considered $P$ 
and $L$ is shown in fig.~\ref{fig1} for both the random ladder- and 
staggered-dimer models. 
Furthermore, the determination of $\overline{M_s}$ is done by extrapolating
the related finite volume staggered structure factors 
to the corresponding bulk results, using a polynomial fit of
the form $a_0 + a_1/L + a_2/L^2 + a_3/L^3$ (The bulk $\overline{M_s}$
is given by $\sqrt{a_0}$). For both the studied models, 
the $L$-dependence of $\overline{S(\pi,\pi)}$ for several considered $P$
is depicted in fig.~\ref{fig2}. The obtained extrapolating values
of $\overline{M_s}$ are shown in fig.~\ref{fig3}.

\subsection{The determination of $\overline{T_N}$}

\begin{figure}
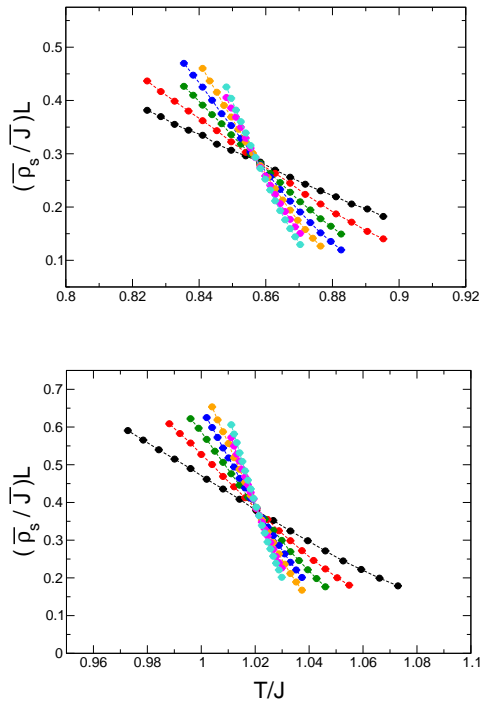

\begin{center}
\vbox{
\includegraphics[width=0.35\textwidth]{ladder_P03_rhos_overJ_L.eps}\vskip0.45cm
\includegraphics[width=0.35\textwidth]{staggered_P005_rhos_overJ_L.eps}
}
\end{center}\vskip-0.5cm
\caption{$(\overline{\rho_s}/\overline{J})L$ as functions of $T/J$ for
$L=$12, 16, 20, 24, 28, 32, and 36 for both the studied models. 
$J$ is 1.0 in our calculations.
While the top panel is for the random ladder-dimer model with $P=0.3$, 
the results of $(\overline{\rho_s}/\overline{J})L$ for the random 
staggered-dimer model with $P=0.05 $ is depicted in the bottom panel. 
The dashed lines are added to guide the eye.}
\label{fig4}
\end{figure}

\begin{figure}
\vskip0.8cm
\begin{center}
\vbox{
\includegraphics[width=0.35\textwidth]{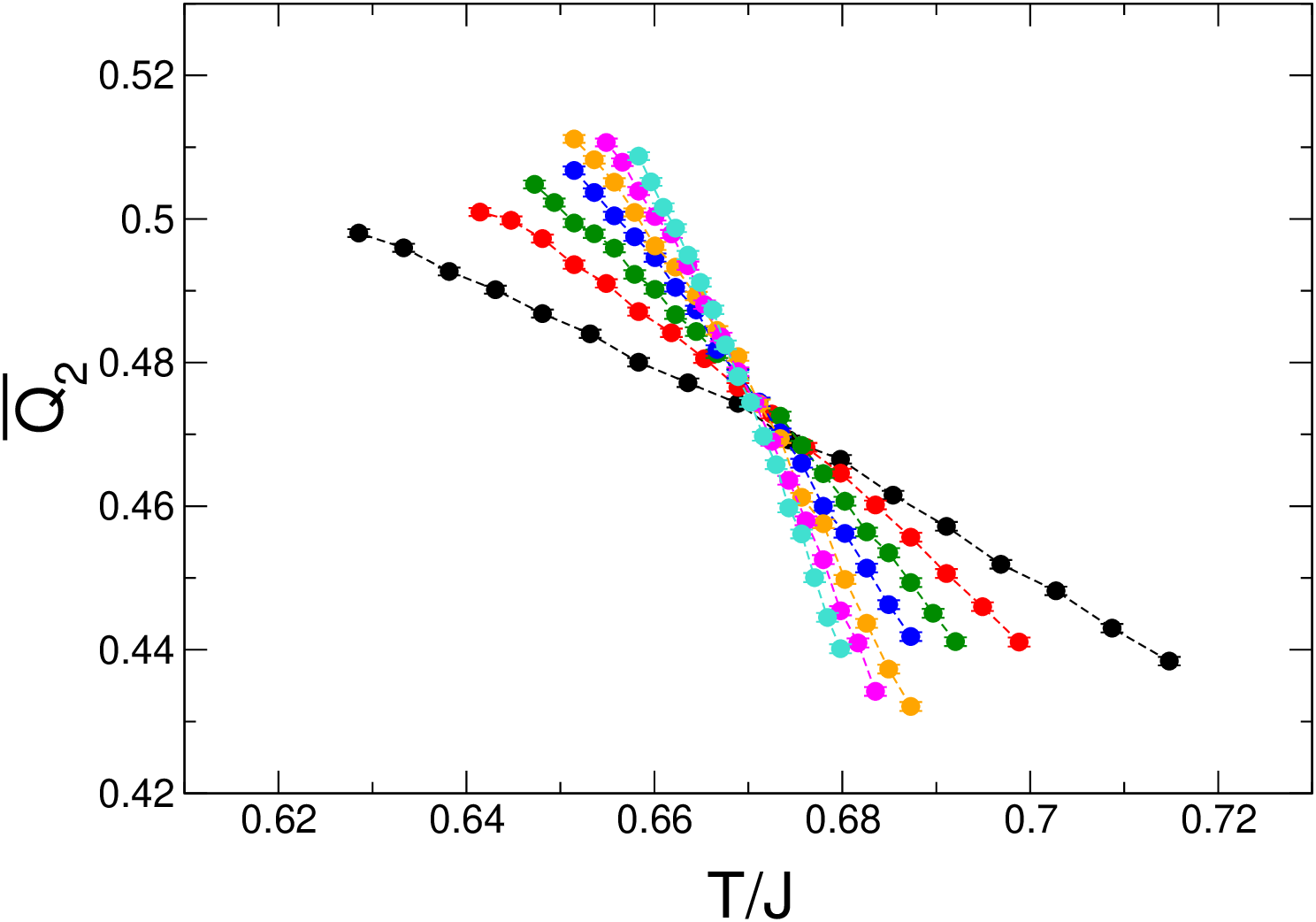}\vskip0.45cm
\includegraphics[width=0.35\textwidth]{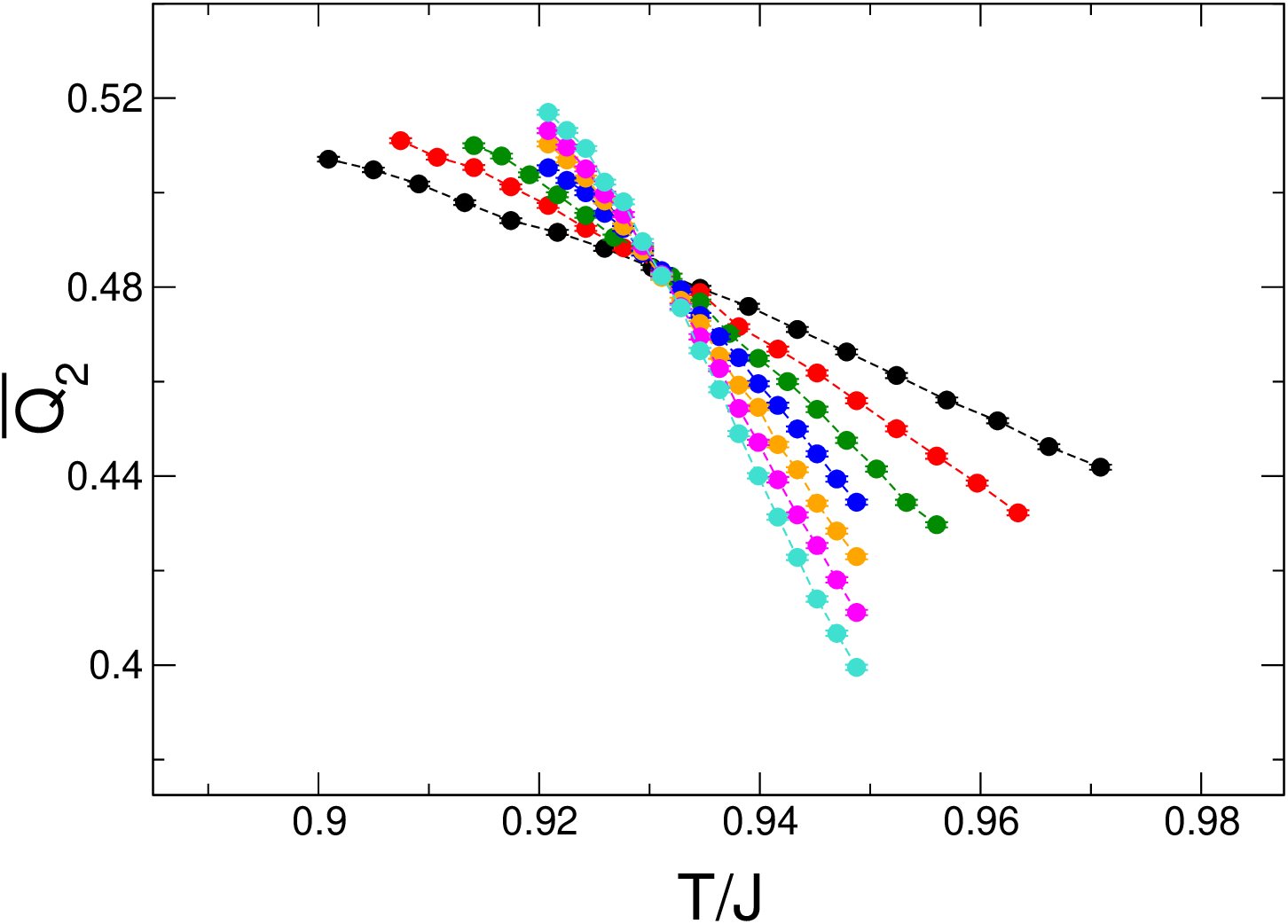}
}
\end{center}\vskip-0.5cm
\caption{$\overline{Q_2}$ as functions of $T/J$ for $L=$12, 16, 20, 24, 28,
32, and 36 for both the studied models. $J$ is 1.0 in our calculations.
While the top panel is for the random ladder-dimer model with $P=0.5$, the 
results of $Q_2$ for the random staggered-dimer model with $P=0.2 $ is 
depicted in the bottom panel. The dashed lines are added to guide the eye.}
\label{fig5}
\end{figure}

\begin{figure}
\vskip0.8cm
\begin{center}
\vbox{
\includegraphics[width=0.35\textwidth]{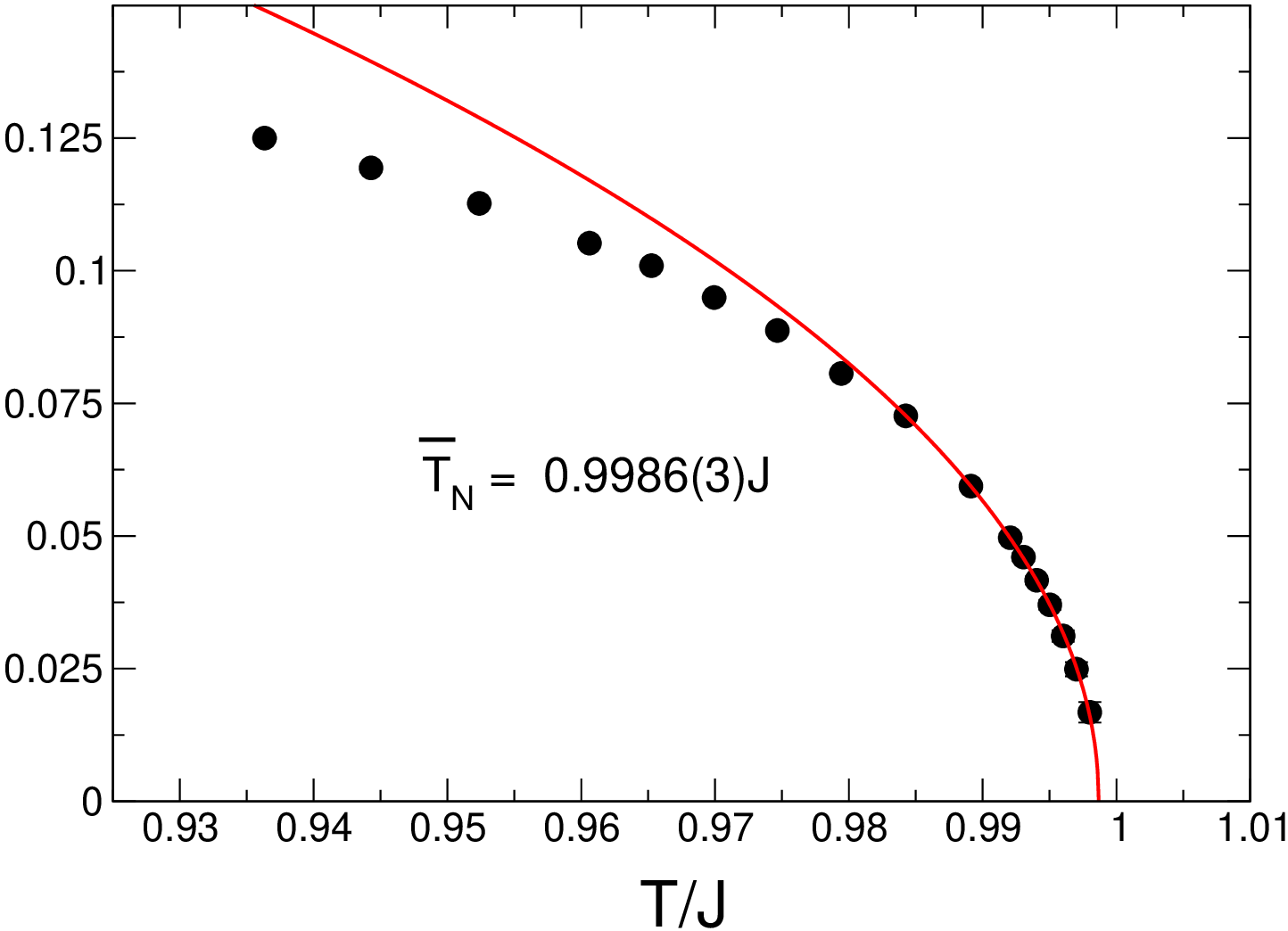}\vskip0.45cm
\includegraphics[width=0.35\textwidth]{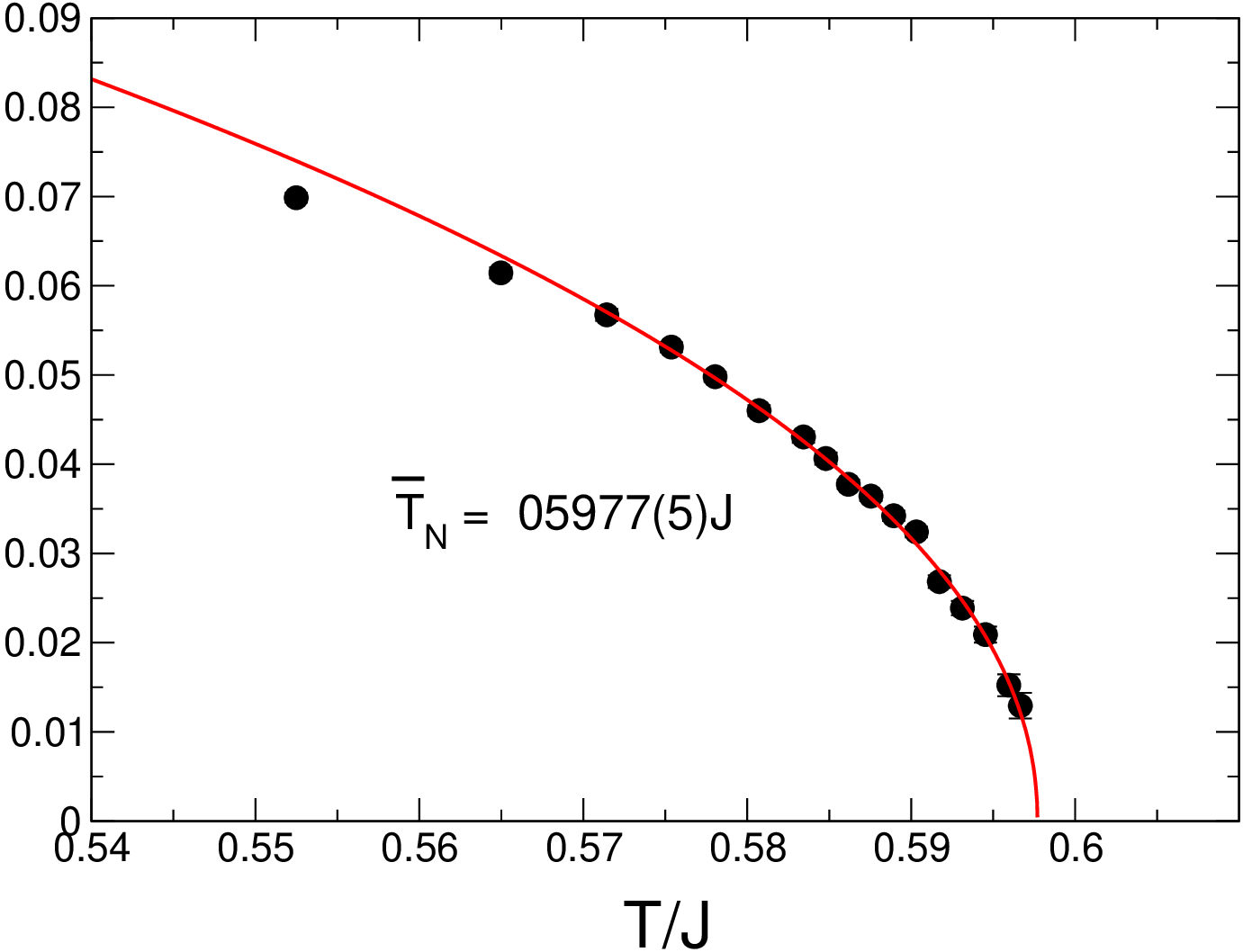}
}
\end{center}\vskip-0.5cm
\caption{$\overline{M_s/}\sqrt{3}$ as functions of $T/J$ for the random 
ladder-dimer model with $P=0.05$ (top panel) and $P=0.55$ (bottom panel). 
$J$ is 1.0 in our calculations. The data points are determined with a up to 
third order polynomial formula in $1/L$ and the quoted errors are estimated 
directly from the fits. The solid lines are obtained using the results of the 
fits (leading singular behaviour ansatz).}
\label{fig5.1}
\end{figure}

\begin{figure}
\vskip0.8cm
\begin{center}
\vbox{
\includegraphics[width=0.35\textwidth]{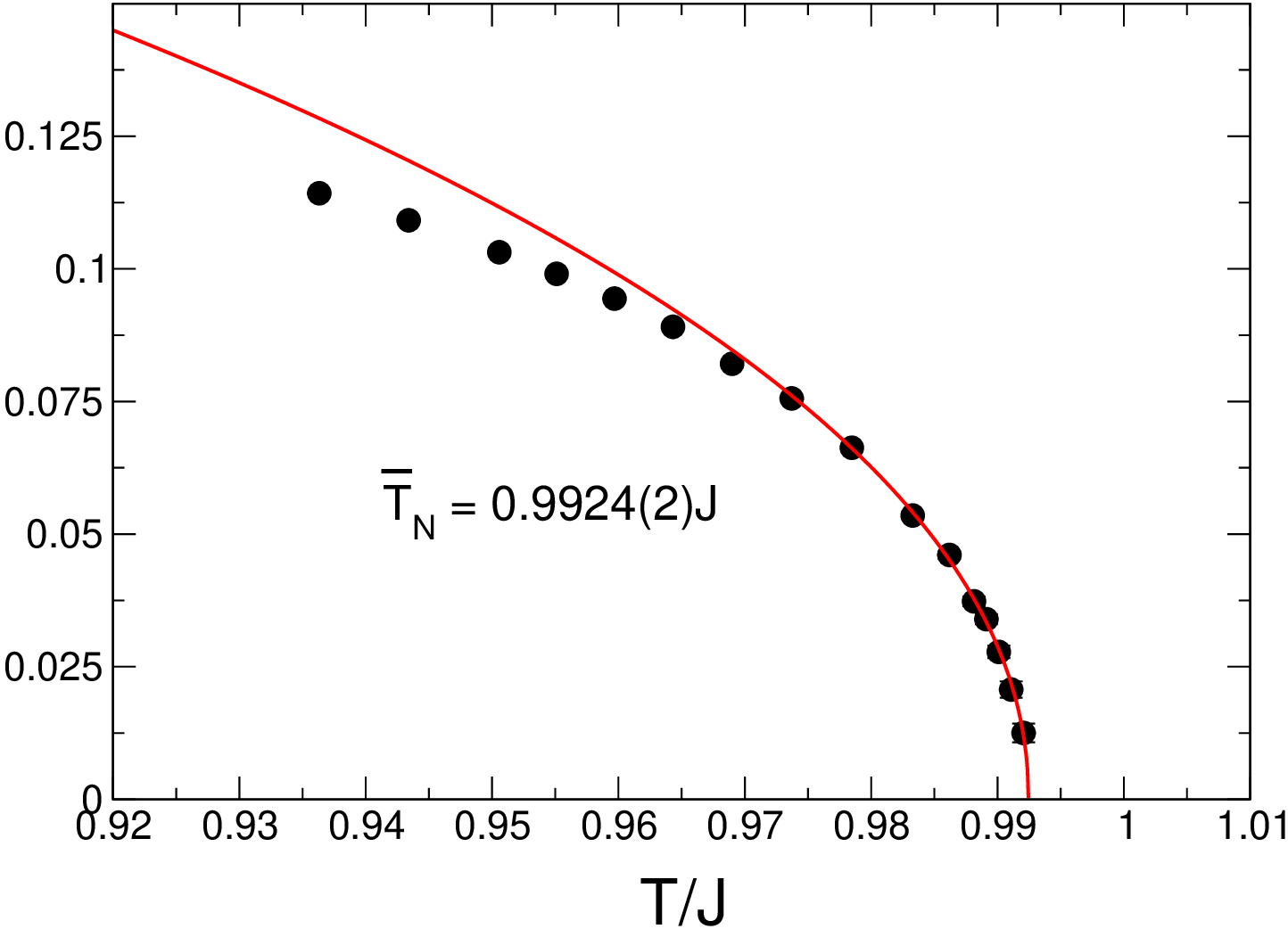}\vskip0.45cm
\includegraphics[width=0.35\textwidth]{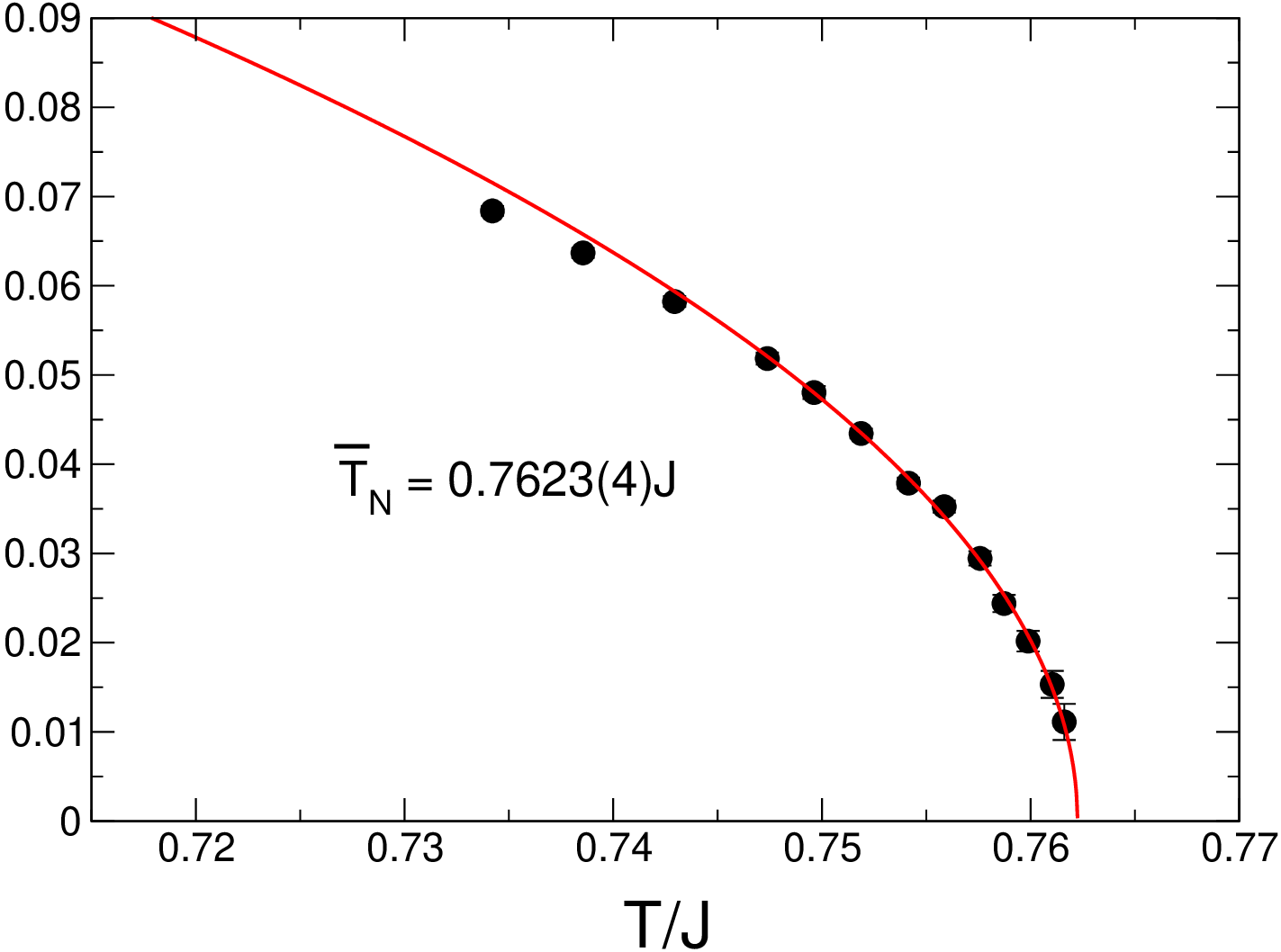}
}
\end{center}\vskip-0.5cm
\caption{$\overline{M_s/}\sqrt{3}$ as functions of $T/J$ for the random 
staggered-dimer model with $P=0.1$ (top panel) and $P=0.4$ (bottom panel). 
$J$ is 1.0 in our calculations. The data points are determined with a up to 
third order polynomial formula in $1/L$ and the quoted errors are estimated 
directly from the fits. The solid lines are obtained using the results of the 
fits (leading singular behaviour ansatz).}
\label{fig5.2}
\end{figure}

The employed observables for calculating $\overline{T_N}$ are 
$(\overline{\rho_s}/\overline{J})L$ as well as $\overline{Q_2}$.
Notice a constraint standard finite-size scaling ansatz of the form 
$(1+b_0L^{-\omega})(b_1 + b_2tL^{1/\nu} +b_3(tL^{1/\nu})^2+...$) 
(or $b_1 + b_2tL^{1/\nu} +b_3(tL^{1/\nu})^2+...$ in some cases), up to fourth 
order in $tL^{1/\nu}$, is used to fit the data. Here $b_i$ for $i=0,1,2,...$ 
are some constants and $t = \frac{T-\overline{T_N}}{\overline{T_N}}$.
It has been demonstrated in Ref.~\cite{Kao14,Kao14.1} that such an analysis 
procedure leads to accurate determination of $\overline{T_N}$. 
The data of $(\overline{\rho_s}/\overline{J})L$ for the random ladder-dimer 
model with $P=0.3$ (top panel) and the random staggered-dimer model with 
$P=0.05 $ (bottom panel) are shown in fig.~\ref{fig4}. The values of 
$\overline{T_N}$ obtained from 
$(\overline{\rho_s}/\overline{J})L$ are listed in table 1. Furthermore,
the quoted errors appearing in table 1 are conservatively estimated, 
based on the standard deviations obtained from a bootstrap method used for the 
fits. Finally, applying a similar analysis to $\overline{Q_2}$ leads to
consistent results of $\overline{T_N}$ with those in table 1. Data points 
of $\overline{Q_2}$ for the random ladder-dimer model with $P = 0.5$ and the 
random staggered-dimer model with $P = 0.2$ are presented in fig.~\ref{fig5}.
In additional to the method of finite-size scaling, $T_N$ can also be 
calculated by studying the singular behaviour of $M_s$ when approaching $T_N$
from $T \le T_N$. We have performed such investigation for the random 
ladder-dimer 
(staggered-dimer) model with $P = 0.05, 0.55$ ($P=0.1, 0.4$). The obtained 
results of $\overline{T_N}$ are in remarkable agreement with those shown in 
table 1. See figs.~\ref{fig5.1} and \ref{fig5.2} for the details.

\begin{table}
\begin{center}
\begin{tabular}{ccccccc}
\hline
$ P $ & & $\overline{T_N}/J$ && $P$ && $\overline{T_N}/J$ \\
\hline
\hline
 0.025 && 1.01030(48) & &0.05 & & 1.02030(53)\\
\hline
 0.05 && 0.99909(52) & &0.1 && 0.99308(48)\\
\hline
 0.1 && 0.97550(40) & &0.2 &&0.93097(49)\\
\hline
 0.2 && 0.92208(40) & & 0.3 &&0.85645(44)\\
\hline
 0.3 && 0.85703(50) & &0.4  && 0.76297(53)\\
\hline
 0.35 && 0.81891(39) & &0.45 & & 0.70586(66)\\
\hline
 0.45 && 0.72636(39) & & 0.5 &&0.63896(56)\\
\hline
 0.5 && 0.66832(44) & &0.55  && 0.55730(21)\\
\hline
 0.55 && 0.59868(62) & &0.6 & & 0.44937(31)\\
\hline
 0.6 && 0.51146(32) & &  &&\\
\hline
 0.625 && 0.45649(28) & &  && \\
\hline
\end{tabular}
\end{center}
\caption{Numerical values of $\overline{T_N}/J$ determined from the 
observable $(\overline{\rho_s}/\overline{J})L$ for various $P$. $J$ is 1.0 in 
our calculations. The first two columns and the last two columns are 
associated with the random ladder- and staggered-dimer models, respectively.}  
\end{table}

\subsection{The determination of $\overline{T^{\star}}$}

\begin{figure}
\vskip0.5cm
\begin{center}
\includegraphics[width=0.375\textwidth]{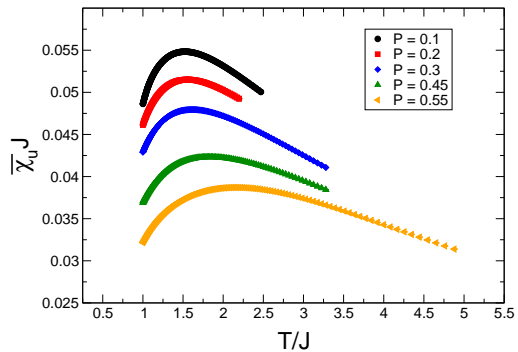}
\end{center}\vskip-0.5cm
\caption{$\overline{\chi_u}$ of the random ladder-dimer model as functions of 
$T/J$ for some values of $P$ and $L=12$. $J$ is 1.0 in our calculations.}
\label{fig6}
\end{figure}

In additional to the universal relation between $T_N/\overline{J}$ and $M_s$,
it is demonstrated in Ref.~\cite{Jin12} that there exists a universal 
connection between the quantities $T_N/T^{\star}$ and $M_s$ as well. Here 
$T^{\star}$ is the temperature where peak of the uniform susceptibility 
$\chi_u$ ($\chi_u = \langle \frac{\beta}{L^3}(\sum_iS^z_i)^2\rangle$),
as a function of the temperatures, takes place.
For both the studied models, the values of $\overline{T^{\star}}$ for the 
considered $P$ are
estimated from the related data of $L=12$ and $L=24$. The numerical values of 
$\overline{\chi_u}$ as functions of $T/J$ for some values of $P$ and $L=12$ are 
shown in fig.~\ref{fig6}.  

\begin{figure}
\vskip0.4cm
\begin{center}
\includegraphics[width=0.365\textwidth]{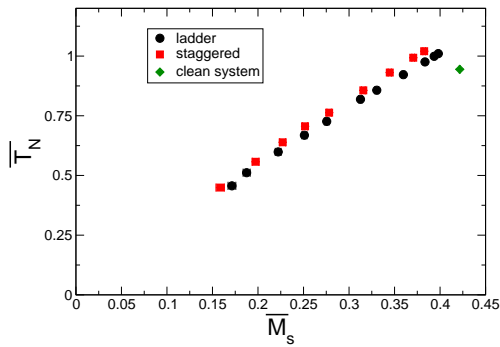}
\end{center}\vskip-0.5cm
\caption{$\overline{T_N}$ as functions of $\overline{M_s}$ for both 
the considered models with the introduced randomness.}
\label{fig7}
\end{figure}

\begin{figure}
\vskip0.4cm
\begin{center}
\includegraphics[width=0.365\textwidth]{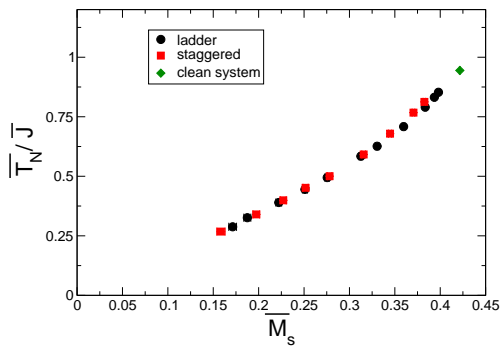}
\end{center}\vskip-0.5cm
\caption{Universal behaviour of $\overline{T_N}/\overline{J}$ 
as functions of $\overline{M_s}$ for the considered models with the introduced
randomness.}
\label{fig8}
\end{figure}

\subsection{The universal relations}

\begin{figure}
\vskip0.4cm
\begin{center}
\includegraphics[width=0.365\textwidth]{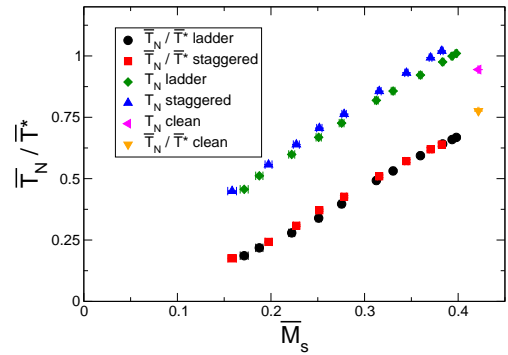}
\end{center}\vskip-0.5cm
\caption{Universal behaviour of $\overline{T_N}/\overline{T^{\star}}$ as 
functions of $\overline{M_s}$ for the considered models with the introduced 
randomness. The corresponding results of $\overline{T_N}$ are shown
in the figure as well. The data points of 
$\overline{T_N}/\overline{T^{\star}}$ contain (around) one percent 
uncertainties due to the errors of $\overline{T^{\star}}$.}
\label{fig9}
\end{figure}

After having obtained $\overline{T_N}$, $\overline{J}$, $\overline{M_s}$, 
and $\overline{T^{\star}}$, as a first step to examine whether universal 
relations, as those established for regular dimerized systems, would appear for
the considered models with the employed randomness, we study $\overline{T_N}$
as functions of $\overline{M_s}$ for both models. The results are shown in 
fig.~\ref{fig7}. As one can see in fig.~\ref{fig7}, no clear signal 
of a universal relation is found in the figure. Remarkably, if the quantity 
$\overline{T_N}/\overline{J}$ is investigated as functions of $\overline{M_s}$,
then indeed the data points of both models fall on top of a universal curve. 
The result is demonstrated in fig.~\ref{fig8}. Similarly, a universal behaviour
between $\overline{T_N}/\overline{T^{\star}}$ and $\overline{M_s}$ is observed as
well (fig.~\ref{fig9}).

\section{Discussions and Conclusions}

Inspired by the universal relations between the N\'eel temperature
$T_N$ and the staggered magnetization density $M_s$ for the clean 3D 
dimerized systems, in this work we study these universal behaviour of 
$T_N$ and $M_s$ for a class of 3D random-exchange quantum Heisenberg models. 
A notable finding here is that these universal properties are even valid for 
the considered models with the introduced randomness. Our results indicate that 
the scope of the validity of these universal properties for 3D quantum 
antiferromagnets is very general. One interesting question is to examine 
whether the data points obtained with other values of $D$ will fall on top of 
the universal curves shown in figs.~\ref{fig8} and \ref{fig9}. Our preliminary 
results related to $\overline{T_N}/\overline{J}$ and $\overline{M_s}$
show that this is not the case. In addition, the data points associated
with clean dimerized ladder model do not form a universal curve with
those data points presented in fig.~\ref{fig8}. As a result, these universal
properties are valid within individual categories, such as the models in
fig.~\ref{fig0} with $D=0.5$ which are investigated in detail here. 
Nevertheless, it is remarkable that the universal relations originally observed 
for the clean models persist even for systems with randomness.

\section*{Acknowledgments}
\vskip-0.25cm
We thank A.~W.~Sandvik for useful discussions.  
This study is partially supported by NCTS (North) and MOST of Taiwan.
The current contact address of F.-J.J. is the Physics Department, 
Duke University, Box 90305, Durham, NC 27708, USA.

\end{document}